\documentclass[journal]{IEEEtran}
\usepackage{graphicx}

\begin{document}

\title{An Efficient Approximation to the Correlated Nakagami-$m$ Sums and its Application in Equal Gain Diversity Receivers}

\author{Nikola Zlatanov, Zoran Hadzi-Velkov, and George K. Karagiannidis 
\thanks{Accepted for IEEE TWireless}
\thanks{N. Zlatanov and Z. Hadzi-Velkov are with the Faculty of Electrical Engineering and Information
Technologies, Ss. Cyril and Methodius University, Skopje, Email:
zoranhv@feit.ukim.edu.mk, nzlatanov@manu.edu.mk}
\thanks{G. K. Karagiannidis is with the Department of Electrical and Computer Engineering, Aristotle University of Thessaloniki, Thessaloniki,
Email: geokarag@auth.gr}
}

\markboth{ }{Shell
\MakeLowercase{\textit{et al.}}: Bare Demo of IEEEtran.cls for
Journals} \maketitle

\begin{abstract}
There are several cases in wireless communications theory where
the statistics of the sum of independent or correlated
Nakagami-$m$ random variables (RVs) is necessary to be known.
However, a closed-form solution to the distribution of this sum
does not exist when the number of constituent RVs exceeds two,
even for the special case of Rayleigh fading. In this paper, we
present an efficient closed-form approximation for the
distribution of the sum of arbitrary correlated Nakagami-$m$
envelopes with identical and integer fading parameters. The
distribution becomes exact for maximal correlation, while the
tightness of the proposed approximation is validated statistically
by using the Chi-square and the Kolmogorov-Smirnov goodness-of-fit
tests. As an application, the approximation is used to study the
performance of equal-gain combining (EGC) systems operating over
arbitrary correlated Nakagami-$m$ fading channels, by utilizing
the available analytical results for the error-rate performance of
an equivalent maximal-ratio combining (MRC) system.
\end{abstract}


\section{Introduction}
\PARstart{T}{he} analytical determination of the the probability
distribution functions (PDF) and the cumulative distribution
functions (CDF) of the sums of independent and correlated signals'
envelopes is rather cumbersome, yielding difficulties in the
theoretical performance analysis of some wireless communications
systems \cite{1}. A closed-form solution for the PDF and the CDF
of the sum of Rayleigh random variables (RVs) has not been
presented for more then 90 years, except when the number of RVs
equals two. The famous Beaulieu series for computing PDF of a sum
of independent RVs were proposed in \cite{2}. Later, a finite
range multifold integral for PDF of the sum of independent and
identically distributed (i.i.d.) Nakagami-$m$ RVs was proposed in
\cite{new5}. A closed-form formula for the PDF of the sum of two
i.i.d. Nakagami-$m$ RVs was given in \cite{new7}-\cite{new7b}.
Exact infinite series representations for the sum of three and
four i.i.d. Nakagami-$m$ RVs was presented in \cite{new8},
although their usefulness is overshadowed by their computational
complexity.

The most famous application, where these sums appear, deals with
the analytical performance evaluation of equal gain combining
(EGC) systems \cite{3}-\cite{9}. Only few papers address the
performance of EGC receivers in correlated fading with
arbitrary-order diversity. In \cite{10}, EGC was studied  by
approximating the moment generating function (MGF) of the output
SNR, where the moments are determined exactly only for
exponentially correlated Nakagami-$m$ channels in terms of
multi-fold infinite series. A completely novel approach for
performance analysis of diversity combiners in equally correlated
fading channels was proposed in \cite{11}, where the equally
correlated Rayleigh fading channels are transformed into a set of
conditionally independent Rician RVs. Based on this technique, the
authors in \cite{12} derived the moments of the EGC output
signal-to-noise ratio (SNR) in equally correlated Nakagami-$m$
channels in terms of the Appell hypergeometric function, and then
used them to evaluate the EGC performance metrics, such as the
outage probability and the error probability (using Gaussian
quadrature with weights and abscissas computed by solving sets of
nonlinear equations).

All of the above works yield to results that are not expressed in
closed form due to the inherent intricacy of the exact sum
statistics. This intricacy can be circumvented by searching for
suitable highly accurate approximations for the PDF of a sum of
arbitrary number of Nakagami-$m$ RVs. Various simple and accurate
approximations to the PDF of sum of independent Rayleigh, Rice and
Nakagami-$m$ RVs had been proposed in \cite{13}-\cite{R4}, which
had been used for analytical EGC performance evaluation. Based on
the ideas given in \cite{1}, the works \cite{15a}-\cite{R4} use
various alternatives of the moment matching method to arrive at
the required approximation.

In this paper, we present a highly accurate closed-form
approximation for the PDF of the sum of non-identical arbitrarily
correlated Nakagami-$m$ RVs with identical (integer) fading
parameters. By applying this approximation, we evaluate the
performance of EGC systems in terms of the known performance of an
equivalent maximal ratio combining (MRC) system \cite{19},
\cite{24}, thus avoiding many complex numerical evaluations
inherent for the methods presented in the aforementioned previous
works for the EGC performance analysis. Although approximate, the
offered closed-form expressions allow to gain insight into system
performance by considering, for example, operation in the low or
high SNR region.

\section{An accurate approximation to the sum of arbitrary correlated Nakagami-$m$ envelopes }
Let $Z$  be a sum of $L$ non-identical correlated Nakagami-$m$
envelopes, $\{Z_k\}_{k=1}^L$, defined as
\begin{equation}\label{1}
Z=\sum_{k=1}^L Z_k \,.
\end{equation}
The envelopes $\{Z_k\}_{k=1}^L$ are distributed according to the
Nakagami-$m$ distribution, whose PDF is given by \cite{1}
\begin{equation}\label{2}
f_{Z_k}(z)=\left(\frac{m_z}{\Omega_k}\right)^{m_z}\frac{2 z^{2
m_z-1}}{\Gamma(m_z)}\exp\left(-\frac{m_z}{\Omega_k}z^2\right),\qquad
z\geq 0 \,,
\end{equation}
with arbitrary average powers $E[Z_k^2]=\Omega_k$, $1\leq k\leq
L$, and the same (integer) fading parameter $m_z$. The power
correlation coefficient between any given pair of envelopes
$(Z_i,Z_j)$ is defined as
\begin{equation}\label{3}
\rho_{ij}=\frac{{\rm{cov}}(Z_i^2,Z_j^2)}{\sqrt{{\rm{var}}(Z_i^2) \
{\rm{var}}(Z_j^2)}} , \qquad i\neq j \,,
\end{equation}
where $E[\cdot]$, $\rm{cov}(\cdot,\cdot)$ and  $\rm{var}(\cdot)$
denote expectation, covariance and variance, respectively.

We propose the unknown PDF of $Z$ be approximated by the PDF of
$R$ defined as
\begin{equation}\label{4}
R=\sqrt{\sum_{k=1}^L R_k^2} \,\,,
\end{equation}
where $R_k$, $1\leq k\leq L$, denote a set of $L$ correlated but identically
distributed  Nakagami-$m$ envelopes with same average
powers, $E[R_k^2]=\Omega_R$, and same fading parameters, $m_R$. The
power correlation coefficients between any given pair $(R_i, R_j)$
is assumed  equal to that of the respective pair of the original envelopes
$(Z_i,Z_j)$, $\rho_{ij}$.

The statistics of $R^2$ is easily seen to be equal to the
statistics of the sum of correlated Gamma RVs. Thus, the MGF of
$R^2$ is represented by [\ref{r24a}, Eq. (11)]
\begin{equation}\label{mgf}
M_{R^2}(s) = \det\left({\bf{I}}-s\frac{\Omega_R}{m_R}{\bf
{\Lambda}}\right)^{-m_R}= \prod_{k=1}^L \left(1 - s
\frac{\Omega_R}{m_R} \lambda_k\right)^{-m_R}
\end{equation}
where ${\bf{I}}$ is the $L \times L$ identity matrix and
${\bf{\Lambda}}$ is the $L \times L$ positive definite matrix
(denoted as the correlation matrix) whose elements are the square
roots of the power correlation coefficients, \vspace{-0.2cm}
\begin{eqnarray} \label{deflambda}
    {\bf{\Lambda}}=\left[
\begin{array}{ccccc}
1&\sqrt{\rho_{12}}&\cdots &\sqrt{\rho_{1L}}\\
\sqrt{\rho_{21}}&1&\cdots &\sqrt{\rho_{2L}}\\
.&.&\cdots &.\\
\sqrt{\rho_{L1}}&\sqrt{\rho_{L2}}&\cdots &1\\
\end{array}
\right] \,.
\end{eqnarray}
The $L$ eigenvalues of the correlation matrix ${\bf{\Lambda}}$ are
denoted by $\lambda_k$, $1\leq k\leq L$.

Throughout literature, the PDF of $R^2$ is determined by using
several different approaches that result in alternative
closed-form solutions, two of which are given by [\ref{r23}, Eq.
(29)] and [\ref{r24}, Eq. (10)]. After a simple RV transformation,
these two alternatives for the PDF of $R$ are expressed as
\begin{equation} \label{7}
f_R(r)= \frac{2r}{\pi}\int_0^\infty\frac{\cos
\left[m_R\sum_{k=0}^{L-1}\arctan\left(t \frac{\Omega_R \lambda_k}
{m_R} \right)-tr^2\right]} {\prod_{k=0}^{L-1}\left[1+\left(t
\frac{\Omega_R \lambda_k} {m_R} \right)^2\right]^{m_R/2}}dt
\end{equation}
\vspace{-7mm}
\begin{eqnarray}
=\frac{2r^{2Lm_R-1}}{\Gamma(Lm_R)}
\left(\frac{m_R}{\Omega_R}\right)^{Lm_R}
\left(\frac{1}{\det({\bf{\Lambda}})}\right)^{m_R} \nonumber \qquad
\qquad \qquad
\end{eqnarray}
\vspace{-3mm}
\begin{eqnarray} \label{7a}
\times
\Phi_2^{(L)}\left(m_R,m_R,\dots,m_R;Lm_R;-\frac{m_R}{\Omega_R}\frac{r^2}{\lambda_1},
\dots ,-\frac{m_R}{\Omega_R}\frac{r^2}{\lambda_L}\right),
\nonumber
\end{eqnarray}
\vspace{-3mm}
\begin{equation}
\end{equation}
where $\Phi_2^{(L)} (\cdot)$ is the confluent Lauricella
hypergeometric function of $L$ variables, defined in \cite{25} and
[\ref{r24}, Eqs. (9)-(10)]. Note that (\ref{7a}) is here presented
to demonstrate existence of an exact closed-form solution, whereas
(\ref{7}) is much more convenient for accurate and efficient
numerical integration. For example, the PDF may be obtained using
the Gauss-Legendre quadrature rule [\ref{r27}, Eq. (25.4.29)] over
(\ref{7}) [\ref{r23}].

Next, we apply the moment matching method to determine the
parameters $\Omega_R$ and $m_R$ of the proposed approximation
(\ref{7})-(\ref{7a}) to the PDF of $Z$. In wireless
communications, moment matching methods are most typically applied
to approximate distributions of the sum of log-normal RVs
\cite{R2}. Most recently, a variant of moment matching, matching
of the normalized first and second moments, had been applied to
arrive at an improved approximation to the sum of independent
Nakagami-$m$ RVs via the $\alpha$-$\mu$ distribution
\cite{R3}-\cite{R4}.

We arrive at required approximation by matching the first and the
second moments of the  powers of $Z$ and $R$, i.e., the second and
fourth moments of the envelopes $Z$ and $R$,
\begin{equation} \label{8}
E[Z^2]=E[R^2], \quad E[Z^4]=E[R^4] \,.
\end{equation}
Matching the first and the second moments of the powers aids the
analytical tractability of the proposed approximation due to the
availability of the MGF of $R^2$ in closed form, given by
(\ref{mgf}). The second and the fourth moments of $R$ are
determined straightforwardly by applying the moment theorem over
(\ref{mgf}), yielding
\begin{eqnarray}
E[R^2]=\frac{d   M_{R^2}(s) }{ds}\Bigg{|}_{s=0}=\Omega_R \sum_{l=1}^L \lambda_l=\Omega_RL \,, \qquad \quad \label{9a}\\
E[R^4]=\frac{d^2 M_{R^2}(s)}
{ds^2}\Bigg{|}_{s=0}=\frac{\Omega_R^2}{m_R}\left[\sum_{l=1}^L\lambda_l^2+
m_R L^2\right] \,. \label{9b}
\end{eqnarray}
Introducing (\ref{9a}) and (\ref{9b})  into (\ref{8}), one obtains
the unknown  parameters for the statistics of $R$
\begin{equation} \label{parametri}
\Omega_R=\frac{E[Z^2]}{L} \,,\quad
m_R=\frac{\sum_{l=1}^L\lambda_l^2}{L^2}\frac{(E[Z^2])^2}{E[Z^4] -
(E[Z^2))^2}\,.
\end{equation}
Using the multinomial theorem and [\ref{r1}, Eq. (137)], the
second and the fourth moments of $Z$ are determined as
\begin{eqnarray}\label{2moment}
E[Z^2]=\sum_{k=1}^L\Omega_k+\frac{2 \Gamma^2(m_z+1/2)}{m_z
\Gamma^2(m_z)} \qquad \qquad \qquad \qquad \nonumber \\
\times \sum_{i=1}^L\sum_{j=i+1}^L\sqrt{\Omega_i\Omega_j} \,
{}_2F_1\left(-1/2,-1/2;m_z;\rho_{ij}\right),
\end{eqnarray}
\begin{eqnarray*}\label{4moment}
E[Z^4] = \frac{m_z+1}{m_z}\sum_{m=1}^L \Omega_m^2 +\frac{6
\Gamma^2(m_z+1)}{m_z^2 \Gamma^2(m_z)} \sum_{i=1}^L \sum_{j=i+1}^L
\Omega_i\Omega_j \qquad \nonumber \\
\times {}_2F_1\left(-1,-1;m_z;\rho_{ij}\right) +\frac{4
\Gamma(m_z+3/2)\Gamma(m_z+1/2)}{m_z^2 \Gamma^2(m_z)}
\qquad \quad \nonumber \\
\times \sum_{i=1}^L
\sum_{j=i+1}^L(\Omega_i^{3/2}\Omega_j^{1/2}+\Omega_i^{1/2}\Omega_j^{3/2})
{}_2F_1\left(-\frac32 ,-\frac12 ;m_z;\rho_{ij}\right) \,\,\,\,
\nonumber\\
+12 \sum_{m=1}^L\sum_{i=m+1}^L \sum_{j=i+1}^L \Omega_m
\sqrt{\Omega_i \Omega_j} E[Z_m^2 Z_i Z_j]
\qquad \qquad \qquad \quad \nonumber \\
+ 12 \sum_{m=1}^L \sum_{i=m+1}^L \sum_{j=i+1}^L \sqrt{\Omega_m}
\Omega_i \sqrt{\Omega_j} E[Z_m Z_i^2 Z_j]
\qquad \qquad \quad \quad \nonumber\\
+12 \sum_{m=1}^L\sum_{i=m+1}^L\sum_{j=i+1}^L \sqrt{\Omega_m
\Omega_i} \Omega_j E[Z_m Z_i Z_j^2] \qquad \qquad \qquad \quad
\end{eqnarray*}
\vspace{-0.3cm}
\begin{equation}
+ \,\,\, 24 \sum_{m=1}^L\sum_{n=m+1}^L\sum_{i=n+1}^L
\sum_{j=i+1}^L\sqrt{\Omega_m \Omega_n \Omega_i \Omega_j} E[Z_m Z_n
Z_i Z_j] \,, \qquad \qquad \qquad \qquad \quad \,\,\,
\end{equation}
where ${}_2F_1(\cdot)$ is the Gauss hypergeometric function
[\ref{r17}]. The joint moments $E[Z_m^2 Z_i Z_j]$, $E[Z_m Z_i^2
Z_j]$, $E[Z_m Z_i Z_j^2]$ and $E[Z_m Z_n Z_i Z_j]$ are not known
in closed-form for arbitrary branch correlation. Exact closed-form
expressions are available only for some particular correlation
models, such as the exponential and the equal correlation models.
For the case or arbitrary correlation, we utilize the method
presented in \cite{26}, where an arbitrary correlation matrix
${\bf{\Lambda}}$ is approximated by its respective Green's matrix,
followed by the application of the available joint moments of the
exponential correlation model.

\begin{table*}[t] \label{Table I}
\caption{Fading parameter $m_R$ for some feasible scenarios with
equal correlation} \centering
\begin{tabular}{c|c|c|c|c|c|c|c|c|c|c|c|c|c|}
& \multicolumn{3}{|c|}{$m_z=1$} &\multicolumn{3}{|c|}{$m_z=2$}
&\multicolumn{3}{|c|}{$m_z=3$}\\
\cline{2-10}
$\rho$ & $L=2$ & $L=3$ &  $L=4$ & $L=2$ & $L=3$ &  $L=4$ &$L=2$ & $L=3$ &  $L=4$  \\
\cline{2-10} & $m_R$ & $m_R$ & $m_R$ &$m_R$ & $m_R$ & $m_R$
& $m_R$ & $m_R$ & $m_R$ \\
\cline{1-10} 0& $0.9552$ & $0.9411$ & $0.9343$  &$1.947$ & $1.93$
& $1.9217$
& $2.943$  &$2.9258$ & $2.9168$ \\
\cline{1-10} 0.2& $0.9195$ & $0.8884$ & $0.8709$  &$1.9102$ &
$1.876$ & $1.8569$
& $2.9068$  &$2.8715$ & $2.8518$ \\
\cline{1-10} 0.4& $0.9156$ & $0.8841$ & $0.8672$  &$1.907$ &
$1.8722$ & $1.8535$
& $2.9039$  &$2.868$ & $2.8487$\\
\cline{1-10} 0.6& $0.9304$ & $0.9056 $& $0.8929$  &$1.9242$ &
$1.8971$ & $1.8831$
& $2.9222$  &$2.8944$ & $2.8799$ \\
\cline{1-10} 0.8& $0.9587$ &$0.9445$ & $0.9374$  &$1.956$ &
$1.9409$ & $1.9333$
& $2.9553$  &$2.9399$ & $2.9321$\\
\cline{1-10}
\end{tabular}
\end{table*}

\begin{table*}[t] \label{Table II}
\caption{Fading parameter $m_R$ for some feasible scenarios with
exponential correlation} \centering
\begin{tabular}{c|c|c|c|c|c|c|c|c|c|c|c|c|c|}
& \multicolumn{3}{|c|}{$m_z=1$} &\multicolumn{3}{|c|}{$m_z=2$}
&\multicolumn{3}{|c|}{$m_z=3$}\\
\cline{2-10}
$\rho$ & $L=2$ & $L=3$ &  $L=4$ & $L=2$ & $L=3$ &  $L=4$ &$L=2$ & $L=3$ &  $L=4$  \\
\cline{2-10} & $m_R$ & $m_R$ & $m_R$ &$m_R$ & $m_R$ & $m_R$
& $m_R$ & $m_R$ & $m_R$ \\
\cline{1-10} 0& $0.9552$ & $0.9411$ & $0.9343$  &$1.947$ & $1.93$
& $1.9217$
& $2.943$  &$2.9258$ & $2.9168$ \\
\cline{1-10} 0.2& $0.9195$ & $0.9033$ & $0.9015$  &$1.9102$ &
$1.892$ & $1.8897$
& $2.9068$  &$2.8878$ & $2.8852$ \\
\cline{1-10} 0.4& $0.9156$ & $0.8887$ & $0.88$  &$1.907$ & $1.877$
& $1.8675$
& $2.9039$  &$2.8728$ & $2.8629$\\
\cline{1-10} 0.6& $0.9304$ & $0.8988 $& $0.8817$  &$1.9242$ &
$1.889$ & $1.87$
& $2.9222$  &$2.8858$ & $2.866$ \\
\cline{1-10} 0.8& $0.9587$ &$0.934$ & $0.9162$  &$1.956$ &
$1.9291$ & $1.9093$
& $2.9553$  &$2.9277$ & $2.9072$\\
\cline{1-10}
\end{tabular}
\end{table*}

\subsection{Equal correlation model}
Equal correlation typically corresponds to the scenario of
multichannel reception from closely spaced diversity antennas
(e.g., three antennas placed on an equilateral triangle). This
model may be employed as a worst case correlation scenario, since
the impact of correlation on system performance for other
correlation models typically will be less severe \cite{19},
\cite{16}.

For this correlation model, the power correlation coefficients are
all equal,
\begin{equation}\label{15}
\rho_{ij}=\rho, \quad i\neq j, \quad 0\leq \rho\leq 1 \,.
\end{equation}
When $m_z$ is assumed to be integer, the unknown joint moments  in
(\ref{4moment}) can be expressed in closed-form as [\ref{r12}, Eq.
(43)]
\begin{eqnarray} \label{15a}
E[Z_m^2 Z_i Z_j] = E[Z_m Z_i^2 Z_j] = E[Z_m Z_i Z_j^2] \qquad
\qquad \nonumber \\ = \left(\frac{1-\sqrt{\rho}}{m_z}\right)^2
W(2,1,1) \,\,, \qquad \qquad
\end{eqnarray}
\begin{eqnarray} \label{15b}
E[Z_m Z_n Z_i Z_j] =
 \left(\frac{1-\sqrt{\rho}}{m_z}\right)^2 W(1,1,1,1) \,, \qquad \quad
\end{eqnarray}
where the coefficients $W(\cdots)$ are determined as
\begin{eqnarray*} \label{W}
W(k_1,...,k_N)
=\left(\prod_{j=1}^N\frac{\Gamma(m_z+k_j/2)}{\Gamma(m_z)}\right)
 \left(\frac{1-\sqrt\rho}{1+(N-1)\sqrt\rho}\right)^{m_z} \nonumber\\
\times
F_A^{(N)}\left(m_z;m_z+\frac{k_1}{2},\cdots,m_z+\frac{k_N}{2};m_z,\cdots,m_z;
\right. \qquad \qquad
\end{eqnarray*}
\vspace{-5mm}
\begin{eqnarray}
\left.
\frac{\sqrt\rho}{1+(N-1)\sqrt\rho},\cdots,\frac{\sqrt\rho}{1+(N-1)\sqrt\rho}\right)\,,
\end{eqnarray}
with $F_A^{(N)}(\cdot)$ denoting the Lauricella $F_A$
hypergeometric function of $N$ variables, defined by [\ref{r17},
Eq. (9.19)] and [\ref{r24b}, Eqs. (11)-(13)].

Note that the coefficient $W(2,1,1)$ needs to be evaluated when $L
\geq 3$, whereas the coefficient $W(1,1,1,1)$ needs to be
evaluated when $L \geq 4$. In Appendix A, $W(2,1,1)$ is reduced to
the more familiar hypergeometric functions, attaining the form
given by (\ref{b7}). $W(1,1,1,1)$ requires numerical evaluation of
the Lauricella $F_A$ function of 4 variables, which can be
computed with desired accuracy by using one of the two numerical
methods presented in [\ref{r24b}, Section IV.A].

The assumption of equal average powers, $\Omega_k = \Omega_Z$, $1
\leq k \leq L$, yields independence of $m_R$ from $\Omega_Z$. For
this case, Table I gives the values of $m_R$ for several
combinations of $\rho$, $L$ and $m_Z$. The use of Table I aids the
practical applicability of our approach for the case of equal
average powers.

For the equal correlation model, the eigenvalues of
${\bf{\Lambda}}$ are exactly found as $\lambda_1 =
1+(L-1)\sqrt\rho$ and $\lambda_k = (1-\sqrt\rho)$ for $2\leq k\leq
L$, so the statistics of  $R^2$ is identical to that of the sum of
a pair of independent Nakagami RVs. Thus, the MGF of $R^2$ is
given by [\ref{r16}, Eq. (9.213)], whereas the PDF of $R$ is given
by [\ref{r16}, Eq. (9.208)]
\begin{eqnarray}\label{23}
f_{R}(r)=\left(\frac{m_R}{\Omega_R}\right)^{m_R L} \qquad \qquad \qquad \qquad \qquad \qquad \qquad \nonumber \\
\times \frac{2r^{2m_R L-1} \exp\left({-m_R
r^2}/{((1-\sqrt{\rho})\Omega_R})\right)} {\Gamma(m_R
L)(1-\sqrt{\rho})^{m_R(L-1)}(1+(L-1)\sqrt{\rho})^{m_R} } \qquad \nonumber\\
\times {}_1F_1\left(m_R; m_R L; \frac{m_R
L\sqrt\rho}{(1-\sqrt{\rho})(1+(L-1)\sqrt{\rho})\Omega_R}r^2\right)
,
\end{eqnarray}
where ${}_1F_1(\cdot)$ is the Kummer confluent hypergeometric
function [\ref{r17}, Eq. (9.210)].

\subsection{Exponential correlation model}
Exponential correlation typically corresponds to the scenario of
multichannel reception from equispaced diversity antennas in which
the correlation between the pairs of combined signals decays as
the spacing between the antennas increases \cite{19}, \cite{16}.

For this correlation model, the power correlation coefficients are
determined as
\begin{equation}\label{12}
\rho_{ij}=\rho^{|i-j|}, \qquad 0\leq \rho\leq 1 \,.
\end{equation}
The unknown joint moments in (\ref{4moment}), $E[Z_m^2 Z_i Z_j]$,
$E[Z_m Z_i^2 Z_j]$, $E[Z_m Z_i Z_j^2]$ and $E[Z_m Z_n Z_i Z_j]$,
can be calculated from [\ref{r10}, Eqs. (11) and (12)]. The
Appendix B derives simpler alternatives to [\ref{r10}, Eqs. (11)
and (12)], which involve a single infinite sum and a familiar
hypergeometric function,
\begin{eqnarray} \label{tri}
E[Z_m^{n_1} Z_i^{n_2} Z_j^{n_3}]=
\frac{|{\bf{\Delta}}|^{m_z}}{\delta_{11}^{m_z+n_1/2}
\delta_{22}^{m_z+n_2/2} \delta_{33}^{m_z+n_3/2}} \qquad \qquad \nonumber\\
\times \, \frac{\Gamma(m_z+n_3/2)}{\Gamma^2(m_z)
m_z^{(n_1+n_2+n_3)/2}} \, \sum_{k=0}^\infty
\left(\frac{\delta_{12}^2}{\delta_{11}
\delta_{22}}\right)^k \qquad \, \nonumber\\
\times \,
\frac{\Gamma(m_z+k+n_1/2)\Gamma(m_z+k+n_2/2)}{\Gamma(m_z+k)k!}
\qquad \quad \nonumber \\
\times {}_2F_1\left(m_z+k+\frac{n_2}{2}, m_z+\frac{n_3}{2}, m_z,
\frac{\delta_{23}^2}{\delta_{22} \delta_{33}}\right),\,
\end{eqnarray}
\begin{eqnarray} \label{cetiri}
E[Z_m Z_n Z_i Z_j]= \frac{|{\bf{\Psi}}|^{m_z}}{(\psi_{11}
\psi_{22} \psi_{33} \psi_{44})^{m_z+1/2}}
\frac{\Gamma^2(m_z+1/2)}{\Gamma^3(m_z) m_z^2} \nonumber \\
\times
\sum_{k=0}^\infty\frac{\Gamma^2(m_z+k+1/2)}{k!\Gamma(m_z+k)}
\left(\frac{\psi_{23}^2}{\psi_{22} \psi_{33}}\right)^k \qquad \,\,\, \nonumber\\
\times \, {}_2F_1\left(m_z+\frac{1}{2}, k+m_z+\frac{1}{2}, m_z,
\frac{\psi_{12}^2}{\psi_{11}\psi_{22}} \right) \nonumber\\
\times {}_2F_1\left(m_z+\frac{1}{2},k+m_z+\frac{1}{2}, m_z,
\frac{\psi_{34}^2}{\psi_{33} \psi_{44}} \right).
\end{eqnarray}
%

In (\ref{tri}), $(n_1,n_2,n_3)$ = $(2,1,1)$ for the calculation of
$E[Z_m^2 Z_i Z_j]$, $(n_1,n_2,n_3)$ = $(1,2,1)$ for the
calculation of $E[Z_m Z_i^2 Z_j]$ and $(n_1,n_2,n_3)$ = $(1,1,2)$
for the calculation of $E[Z_m Z_i Z_j^2]$. The matrix ${\bf{\Delta}} =
[\delta_{i,j}]$ is the inverse of ${\bf{\Lambda}}$'s principal
submatrix composed of the $m$-th, $i$-th and $j$-th
rows and columns of ${\bf{\Lambda}}$, whereas the matrix
${\bf{\Psi}} = [\psi_{i,j}]$ is the inverse of ${\bf{\Lambda}}$'s
principal submatrix composed of the $m$-th, $n$-th,
$i$-th and $j$-th rows and columns of ${\bf{\Lambda}}$,
\begin{eqnarray*}
 \label{Delta_definition}
    {\bf{\Delta}}=\left[
\begin{array}{ccccc}
1&\sqrt{\rho_{mi}}&\sqrt{\rho_{mj}}\\
\sqrt{\rho_{im}}&1&\sqrt{\rho_{ij}}\\
\sqrt{\rho_{jm}}&\sqrt{\rho_{ji}}&1\\
\end{array}
\right]^{-1},
\end{eqnarray*}
\begin{eqnarray}\label{Psi_definition}
    {\bf{\Psi}}=\left[
\begin{array}{ccccc}
1&\sqrt{\rho_{mn}}&\sqrt{\rho_{mi}}&\sqrt{\rho_{mj}}\\
\sqrt{\rho_{nm}}&1&\sqrt{\rho_{ni}}&\sqrt{\rho_{nj}}\\
\sqrt{\rho_{im}}&\sqrt{\rho_{in}}&1&\sqrt{\rho_{ij}}\\
\sqrt{\rho_{jm}}&\sqrt{\rho_{jm}}&\sqrt{\rho_{ji}}&1\\
\end{array}
\right]^{-1} \, .
\end{eqnarray}
The exactness of (\ref{tri})-(\ref{cetiri}) arise from the fact
that both matrices ${\bf{\Delta}}$ and ${\bf{\Psi}}$ are
tridiagonal matrices due to (\ref{12}) [\ref{r26}, Section IV].
Introducing (\ref{tri})-(\ref{cetiri}) into (\ref{4moment}), one
obtains the closed-form expression for $E[Z^4]$, which is omitted
here for brevity. Combining (\ref{2moment})-(\ref{4moment}) into
(\ref{parametri}), one obtains the unknown parameters $\Omega_R$
and $m_R$ for the statistics of $R$.

The assumption of equal average powers, $\Omega_k = \Omega_Z$, $1
\leq k \leq L$, again renders independence of $m_R$ from
$\Omega_Z$ for the exponential correlation model. Under such
assumptions, Table II displays the values of $m_R$ for several
illustrative combinations of $\rho$, $L$ and $m_Z$.

\subsection{Arbitrary correlation model}
In the general case of arbitrary branch correlations, the
correlation matrix ${\bf{\Lambda}}$ is approximated by its
appropriate Green's matrix, ${\bf{C}}$, utilizing the method
presented in [\ref{r26}, Section IV]. Since principal submatrices
of Green's matrices are also Green's matrices, the matrices
${\bf{\Delta}}$ and ${\bf{\Psi}}$, defined by
(\ref{Delta_definition}), are determined to be tridiagonal,
yielding direct applicability of the results presented in Section
II.B to determine the unknown parameters $\Omega_R$ and $m_R$ for
the statistics of $R$. Thus, the statistics of $Z$ are
approximated by the statistics of $R$, whose arbitrary correlation
matrix ${\bf{\Lambda}}$ is approximated by the Green's matrix
${\bf{C}}$. In the following subsection, we illustrate the highly
accurate approximation to the PDF of $Z$ facilitated by this
approach.

\begin{table*}[t] \label{Table III}
\caption{\hspace{2cm} Significance levels of C-S and K-S tests for
goodness \newline of fit  between the exact and the approximative
distributions of Fig. 1} \centering
\begin{tabular}{c|c|c|c|c|c|c|c|c|c|}
& \multicolumn{4}{|c|}{$m_z=1$} &\multicolumn{4}{|c|}{$m_z=3$}\\
\cline{2-9}
$\rho$& \multicolumn{2}{|c|}{$L=2$} &\multicolumn{2}{|c|}{$L=5$} & \multicolumn{2}{|c|}{$L=2$} &\multicolumn{2}{|c|}{$L=5$}\\
\cline{2-9} & $\alpha_{CS}$ & $\alpha_{KS}$ & $\alpha_{CS}$
&$\alpha_{KS}$
&$\alpha_{CS}$ & $\alpha_{KS}$ & $\alpha_{CS}$  &$\alpha_{KS}$  \\
\cline{1-9} 0.2 & $0.004$ & $0.02$ & $0.17$  & $0.06 $ & $<0.001$
& $<0.001$ & $<0.001$   &$<0.001$   \\
\cline{1-9}
0.7 & $0.04$ & $0.03$ & $0.2$  & $0.18$ & $<0.001$  & $<0.001$ & $<0.001$   &$<0.001$  \\
\cline{1-9}
\end{tabular}
\end{table*}

\begin{table*} [t] \label{Table IV}
\caption{\hspace{2cm} Significance levels of C-S and K-S tests for
goodness \newline of fit between the exact and the approximative
distributions of Fig. 2} \centering
\begin{tabular}{c|c|c|c|c|c|c|c|c|c|}
& \multicolumn{4}{|c|}{$m_z=1$} &\multicolumn{4}{|c|}{$m_z=3$}\\
\cline{2-9}
$\rho$& \multicolumn{2}{|c|}{$L=2$} &\multicolumn{2}{|c|}{$L=5$} & \multicolumn{2}{|c|}{$L=2$} &\multicolumn{2}{|c|}{$L=5$}\\
\cline{2-9} & $\alpha_{CS}$ & $\alpha_{KS}$ & $\alpha_{CS}$
&$\alpha_{KS}$
&$\alpha_{CS}$ & $\alpha_{KS}$ & $\alpha_{CS}$  &$\alpha_{KS}$  \\
\cline{1-9} 0.2& $<0.001$ & $0.02$ & $<0.001$ & $<0.001$& $<0.001$
& $<0.001$ &$<0.001$ &$<0.001$   \\
\cline{1-9}
0.7 &$0.08$&  $0.04$ & $0.02$ & $0.17$ & $<0.001$ & $<0.001$ &$<0.001$ &$<0.001$  \\
\cline{1-9}
\end{tabular}
\end{table*}

\begin{figure}[!ht]
\begin{center}
\includegraphics[width=3.7in]{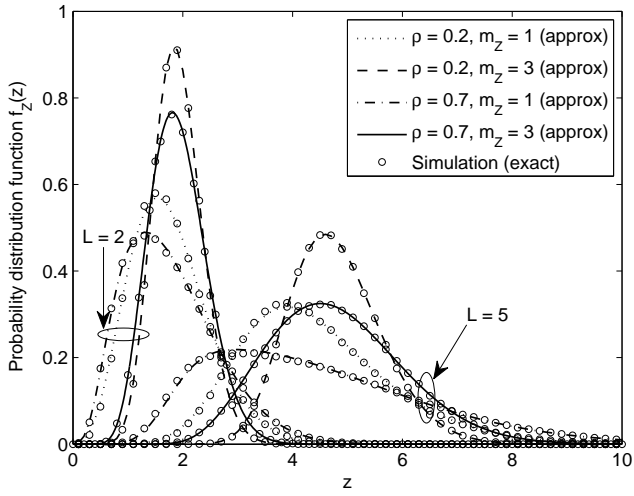}
\end{center}
\vspace{-5mm}
\begin{center}
\caption{Exact (obtained by simulation) and the approximative
analytical PDFs to the sum of equally correlated Nakagami-$m$ RVs
with equal average powers, when $\Omega_Z = 1$}\label{sl4}
\end{center}
\end{figure}

\subsection{Validation via statistical goodness-of-fit tests} We
now statistically validate the proposed PDF approximations for
equal, exponential and arbitrary branch correlation by using two
different goodness-of-fit tests. The Chi-square (C-S) and
Kolmogorov-Smirnov (K-S) tests provide two different statistical
metrics, $\chi_n^2$ and $D_n$, which describe the discrepancy
between the observed samples of $Z$ and the samples expected under
the analytical distribution (\ref{7})-(\ref{7a}).

Each metric is averaged over 100 statistical samples, where each
statistical sample comprises of 10000 independent random samples
of $Z$. The random samples of $Z$ are generated by computer
simulations of correlated Nakagami-$m$ RVs based on the method
proposed in [\ref{r21}, Section VII].

For each metric, we calculate the significance level $\alpha$
from the C-S and K-S distributions, respectively denoted as $\alpha_{CS}$
and $\alpha_{KS}$. The significance level $\alpha$
represents the probability of rejecting the tested null hypothesis
($H_0$: "the random samples of $Z$, obtained from (\ref{1}),
belong to the distribution given by (\ref{7})-(\ref{7a})"), when it is
actually true. The small values of $\alpha$ indicate a good fit.

Note that, significance levels $\alpha$ less then $0.2$ still
indicate a good fit, due to the rigourousness of both C-S and K-S
tests in accepting the null hypothesis $H_0$.

\begin{figure}[!ht]
\begin{center}
\includegraphics[width=3.7in]{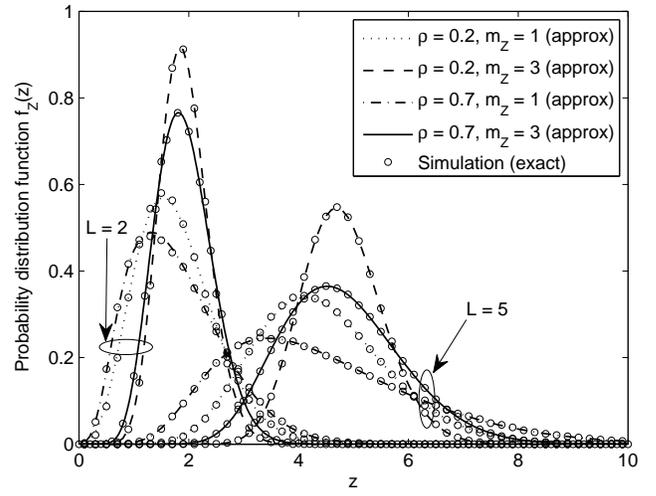}
\end{center}
\vspace{-3mm} \caption{Exact (obtained by simulation) and the
approximative analytical PDFs to the sum of exponentially
correlated Nakagami-$m$ RVs with equal average powers, when
$\Omega_Z = 1$}\label{sl3}
\end{figure}

\subsubsection{Equal and exponential correlation}
For the equal and exponential correlation models, the
goodness-of-fit testing is conducted for combinations of the
followings input parameters: $L = 2$ and $5$, $m_Z = 1$ and $3$,
$\rho = 0.2$ and $0.7$, whereas the average powers of $Z_k$ are
assumed equal to unity ($\Omega_Z = 1$). The needed fading
parameter $m_R$ of distribution (\ref{7})-(\ref{7a}) is obtained
directly from Tables I and II, whereas the average power
$\Omega_R$ is calculated from (\ref{parametri}).

Figs. 1 and 2 depict the excellent (visual) match between the
histogram obtained from generated samples of $Z$ and the proposed
approximation, for the cases of equal and exponential correlation
models, respectively. Tables III and IV complement Figs. 1 and 2,
by presenting the significance levels $\alpha$ for the
corresponding input parameters' combinations. The Table III and
the Table IV entries reveal the very low significance levels
$\alpha$ for all input parameters' combinations, thus proving an
excellent goodness of fit in statistical sense.

\subsubsection{Arbitrary correlation}
For illustrative purposes, we use same two example correlation
matrices from [\ref{r26}, Sections V.B and V.D],
$\Sigma_{3\_\rm{lin}}$ and $\Sigma_{4\_\rm{circ}}$, here denoted
as ${\bf{\Lambda_1}}$ and ${\bf{\Lambda_2}}$, respectively. They
are approximated by their Green's matrices $C_{3\_\rm{lin}}$ and
$C_{4\_\rm{circ}}$, here denoted as ${\bf{C_1}}$ and ${\bf{C_2}}$,
respectively.

Using ${\bf{C_1}}$ and ${\bf{C_2}}$, one obtains the needed
tridiagonal matrices ${\bf{\Delta}}$ and ${\bf{\Psi}}$ from their
definitions given by (\ref{Delta_definition}). The required joint
moments are then calculated from (\ref{tri}) and (\ref{cetiri}),
which are then substituted into (\ref{2moment}) and
(\ref{4moment}) to calculate $E[Z^2]$ and $E[Z^4]$, and then
(\ref{parametri}) is used to describe the statistics of $R$.

\begin{table}[t] \label{Table V}
\caption{Significance levels of C-S and K-S tests for goodness of
fit between exact and approximative distributions of Fig. 3}
\centering
\begin{tabular}{c|c|c|c|c|c|c|c|c|c|}
& \multicolumn{2}{|c|}{$m_z=1$} &\multicolumn{2}{|c|}{$m_z=3$}\\
\cline{2-5}
& $\alpha_{CS}$ & $\alpha_{KS}$ & $\alpha_{CS}$  &$\alpha_{KS}$   \\
\cline{1-5} $\Lambda_1$& $0.19$ & $0.18 $ & $<0.001$ &$<0.001$   \\
\cline{1-5}
$\Lambda_2$& $0.12$ & $0.11$ & $<0.001$  &$<0.001$  \\
\cline{1-5}
\end{tabular}
\end{table}

\begin{figure}[!ht]
\begin{center}
\includegraphics[width=3.7in]{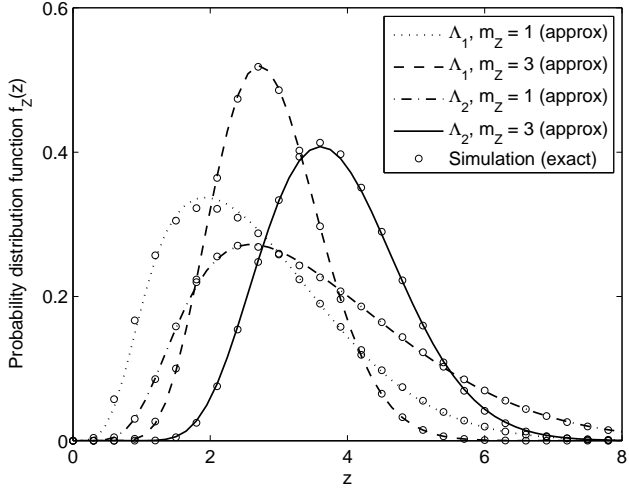}
\vspace{-5mm}
\end{center}
\caption{Exact (obtained by simulation) and the approximative
analytical PDFs to the sum of correlated Nakagami-$m$ RVs with
equal average powers, when $\Omega_Z = 1$), and correlation
matrices ${\bf{\Lambda_1}}$ and ${\bf{\Lambda_2}}$}\label{sl1}
\end{figure}

Fig. 3 depicts the excellent (visual) match between the histogram
obtained from generated samples of $Z$ and the proposed
approximation (\ref{7})-(\ref{7a}), for the two example
correlation matrices $\Lambda_1$ and $\Lambda_2$. Table V
complements Figs. 3, by revealing the very low significance levels
$\alpha_s$, thus again proving an excellent goodness of fit.

\subsection{Validation in case of maximal correlation}
We now consider the case of maximal correlation coefficient
between any pair of Nakagami-$m$ envelopes $Z_i$ and $Z_j$, i.e.,
$\rho_{ij} = 1$. It indicates a perfect linear relationship
between these pairs, which, after applying the model from
[\ref{r12}, Eq. (37)], can be defined as $Z_i = \sqrt{\Omega_i} \,
Z_0$ for $1 \leq i \leq L$, where $Z_0$ is an auxiliary
Nakagami-$m$ RV with unity average power and same fading parameter
$m_z$. After replacing the latter expression into (\ref{1}), $Z$
is transformed into a Nakagami-$m$ RV with fading parameter $m_z$
and average power of
\begin{equation}\label{maxcorr1}
E[Z^2]=\left(\sum_{i=1}^L \sqrt{\Omega_i}\right)^2 = \sum_{i=1}^L
\sum_{j=1}^L \sqrt{\Omega_i \Omega_j} \,,
\end{equation}
which agrees with (\ref{2moment}) when $\rho_{ij} \rightarrow 1$.

Replacing $\rho_{ij} = 1$ into (\ref{deflambda}), the $L-1$
eigenvalues of the matrix $\bf{\Lambda}$ turn up equal to 0,
except $\lambda_1 = L$. After plugging these eigenvalues into
(\ref{mgf}), $R$ is transformed into a Nakagami-$m$ RV with fading
parameter $m_R$ and average power $L \Omega_R$. After the moment
matching, $\Omega_R$ and $m_R$ can be obtained from
(\ref{parametri}), as
\begin{equation}
\Omega_R=\frac{1}{L} \sum_{i=1}^L \sum_{j=1}^L \sqrt{\Omega_i
\Omega_j},\quad m_R=\frac{(E[Z^2])^2}{E[Z^4]-(E[Z^2])^2} = m_z \,,
\end{equation}
respectively, where the latter equality is attributed to the
definition of the Nakagami-$m$ fading parameter, given by
[\ref{r1}, Eq. (4)].

Thus, maximal correlation yields (\ref{7})-(\ref{7a}) as an
accurate distribution of $Z$, when our moment matching approach is
applied. This conclusion further validates our approach.
\vspace{-5mm}
\section{Application to the performance analysis of EGC receivers}
We now consider a typical $L$-branch EGC diversity receiver
exposed to slow and flat Nakagami-$m$ fading. The envelopes of the
branch signals $Z_k$ are non-identical   correlated Nakagami-$m$
random processes with PDFs given by (\ref{2}), whereas their
respective phases are i.i.d. uniform random processes. Each branch
is also corrupted by additive white Gaussian noise (AWGN) with
power spectral density $N_0/2$, which is added to the useful
branch signal. In the EGC receiver, the random phases of the
branch signals are compensated (co-phased), equally weighted and
then summed together to produce the decision variable. The
envelope of the composite useful signal, denoted by $Z$, is given
by (\ref{1}), whereas the composite noise power is given by
$\sigma_{EGC}^2=LN_0/2$, resulting in the instantaneous output SNR
given by
\begin{equation}\label{16}
\gamma_{EGC}=\frac{Z^2}{2\sigma_{EGC}^2}=\frac{1}{LN_0}\left(\sum_{k=1}^L
Z_k\right)^2=\left(\sum_{k=1}^L G_k\right)^2
\end{equation}
where RVs $G_k=Z_k/\sqrt{LN_0}$, $1\leq k\leq L$, form a set of
$L$ non-identical equally correlated Nakagami-$m$ RVs with
$E[G_k^2] =\bar\gamma_k/L$, same fading parameters $m_Z$ and
correlation coefficient $\rho_{ij}$ between branch pair $(i,j)$.
Note that $\bar\gamma_k=\Omega_k/N_0$  denotes the average SNR in
$k$-th branch.

Using the results from Section II, the MGF and the PDF of
(\ref{16}) can be approximated using (\ref{mgf}) and
(\ref{7})-(\ref{7a}), respectively, when $\Omega_R$ is replaced by
$\bar\gamma_R=\Omega_R/(LN_0)$. These approximations are then used
to determine the outage probability $F_{\gamma_{EGC}}$ and the
error probability $\bar P_{EGC}$ of an $L$-branch EGC systems in
correlated Nakagami-$m$ fading with high accuracy.

\begin{figure}[!ht]
\begin{center}
\includegraphics[width=3.7in]{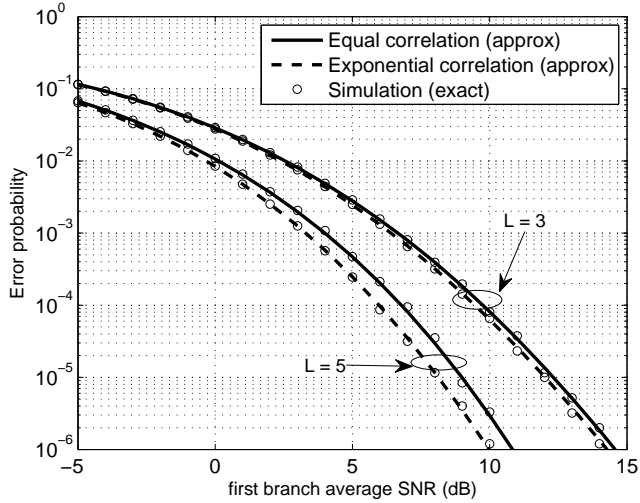}
\vspace{-5mm}
\end{center}
\caption{Exact and approximate error probabilities of an EGC
receiver with correlated Nakagami-$m$ branches, when $m_z = 2$,
$\mu = 0$ and $\rho=0.7$}\label{sl5}
\end{figure}

\subsection{Outage probability}
The outage probability of the EGC system with arbitrary correlated
Nakagami-$m$ fading branches, whose output SNR drops below
threshold $t$, is approximated by the known outage probability
expressions of an equivalent MRC system [\ref{r23}, Eq. (28)],
[\ref{r24}, Eq. (13)],
\begin{eqnarray*}\label{op1}
F_{\gamma_{EGC}}(t) \approx F_{\gamma_{MRC}}(t) \qquad \qquad \qquad \qquad \qquad \qquad \qquad \nonumber \\
=\frac{1}{2}-\frac{1}{\pi}\int_0^\infty\frac{\sin
\left[m_R\sum_{k=0}^{L-1}\arctan \left(x \frac{\Omega_R \lambda_k}
{m_R}\right)-xt\right]}
{\prod_{k=0}^{L-1}\left[1+\left(x \frac{\Omega_R \lambda_k} {m_R}\right)^2\right]^{m_R/2}}\frac{dx}{x}\nonumber\\
\end{eqnarray*}
\begin{eqnarray}
=\frac{1}{\Gamma(1+Lm_R)}\left(\frac{m_R}{\Omega_R}t\right)^{L m_R}
\frac{1}{\det(\Lambda)} \qquad \qquad \qquad \qquad \nonumber \\
\times \Phi_2^{(L)}\Big(m_R,m_R,\dots,m_R;1+Lm_R; \qquad \qquad \qquad \nonumber \\
\left.
-\frac{m_R}{\Omega_R\lambda_1}t,-\frac{m_R}{\Omega_R\lambda_2}t,\dots,
-\frac{m_R}{\Omega_R\lambda_L}t\right) \,.
\end{eqnarray}
For the equal correlation model, (\ref{op1}) can be simplified
using [\ref{r17}, Eq. (2.1.3(1))].

\subsection{Average error probability}
The average error probability of the correlated Nakagami-$m$ EGC
system with BPSK modulation / coherent demodulation is
approximated using the available expressions for the average error
probability of the equivalent MRC systems. Based on [\ref{r16},
Eq. (9.11)] and [\ref{r24}, Eq. (17)], the error performance of
this EGC system is alternatively approximated as
\begin{equation}
\bar P_{EGC-BPSK} \approx \bar P_{MRC-BPSK} = \frac{1}{\pi}
\int_0^{\pi/2} M_{R^2}\left(\frac{-1}{\sin^2\theta}\right) d\theta
\label{ber1}
\end{equation}
\begin{eqnarray}
=\frac{\Gamma(Lm_R+1/2)}{2\sqrt{\pi}\Gamma(Lm_R+1)}
\left(\frac{m_R}{\Omega_R} \right)^{Lm_R}
\left(\frac{1}{\det(\Lambda)}\right)^{m_R} \qquad \quad \nonumber\\
\times \, F_D^{(L)} \Big( Lm_R+1/2,m_R,\dots,m_R;Lm_R+1; \nonumber\\
\left. -\frac{m_R}{\Omega_R\lambda_1},
-\frac{m_R}{\Omega_R\lambda_2},\dots,-\frac{m_R}{\Omega_R\lambda_L}\right)
\,. \label{ber2}
\end{eqnarray}

\begin{figure}[!ht]
\begin{center}
\includegraphics[width=3.7in]{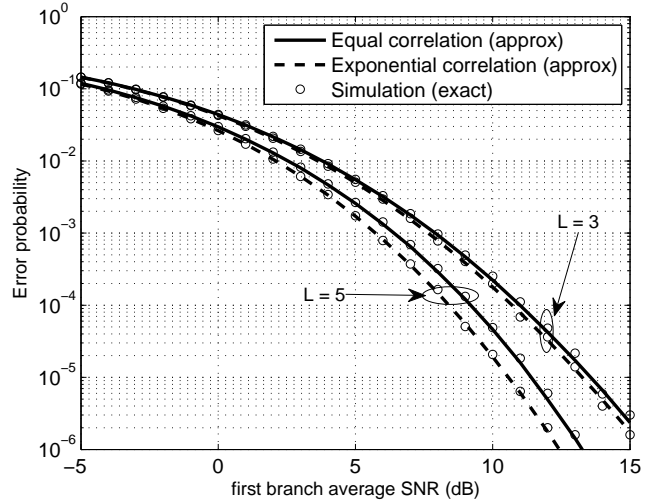}
\vspace{-5mm}
\end{center}
\caption{Exact and approximate error performance of an EGC
receiver with correlated Nakagami-$m$ branches, when $m_z = 2$,
$\mu = 0.3$ and $\rho=0.7$}\label{sl6}
\end{figure}

\begin{figure}[!ht]
\begin{center}
\includegraphics[width=3.7in]{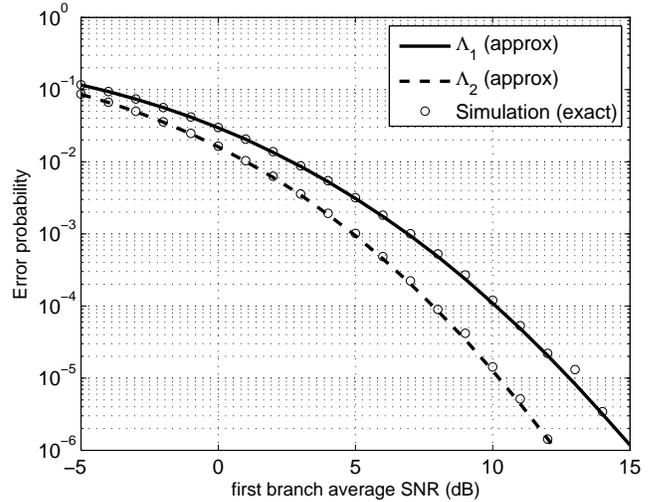}
\vspace{-5mm}
\end{center}
\caption{Exact and approximate error performance of an EGC
receiver with correlated Nakagami-$m$ branches, when correlation
is described by correlation matrices ${\bf{\Lambda_1}}$ and
${\bf{\Lambda_2}}$, $m_z = 2$ and $\mu = 0$}\label{sl2}
\end{figure}

\noindent In (\ref{ber1}), $M_{R^2}(\cdot)$ is replaced with the
MGF given by (\ref{mgf}). In (\ref{ber2}), $F_D^{(L)}(\cdot)$
denotes the Lauricella $F_D$ hypergeometric function of $L$
variables, defined in \cite{25} and [\ref{r24}, Eq. (18)]. For the
equal correlation model, the average error probability can be
calculated using [\ref{r19}, Eq. (32)], which is a special case of
(\ref{ber2}).

Note that (\ref{ber2}) is here presented to demonstrate existence
of an exact closed-form solution, whereas (\ref{ber1}) is much
more convenient for accurate and efficient numerical integration.
For example, the average error probability may be obtained by
applying the Gauss-Chebyshev quadrature rule [\ref{r27}, Eq.
(25.4.38)] over (\ref{ber1}). In the case of the balanced
diversity branches with equal or exponential correlation, the
combination of this quadrature rule with Tables I and II allows
efficient and extremely accurate evaluation of the EGC
performance.

The average error probability of correlated Nakagami-$m$ EGC
system with BFSK modulation / non-coherent demodulation is
approximated by known expression of the equivalent MRC system
[\ref{r24a}, Eq. (16)], $\bar P_{EGC-BFSK} \approx \bar
P_{MRC-BFSK}= \frac12 \, M_{R^2} (-\frac12 )$, where
$M_{R^2}(\cdot)$ is given by (\ref{mgf}).

\subsection{Validation via Monte-Carlo simulations}
Next, we illustrate the tightness of the error performance of an
correlated Nakagami-$m$ EGC system with BPSK modulation / coherent
demodulation to that of the equivalent MRC system. The results for
the actual EGC system are obtained by Monte-Carlo simulations,
whereas those of the equivalent MRC system are obtained using
(\ref{ber1}).

\subsubsection{Equal and exponential branch correlation}
Figs. 4 and 5 displays the comparative error performance of the
actual EGC and the equivalent MRC systems, for several
combinations of $(\rho, L, m_Z, \Omega_k)$. In order to
accommodate unequal average branch powers (thus, unequal average
branch SNRs), we used the exponentially decaying profile, modelled
as
\begin{equation}
\Omega_k = \Omega_1 \exp(-\mu (k-1)), \quad 1\leq k\leq L \,,
\end{equation}
where $\Omega_1$ is the average power of branch 1 and $\mu$ is the
decaying exponent, with $\mu = 0$ denoting the case of branches
with equal power (i.e., the balanced branches).

\subsubsection{Arbitrary branch correlation}
Fig. 6 depicts the comparative error performances of the EGC with
same correlation matrices from Section II.D, ${\bf{\Lambda_1}}$
and ${\bf{\Lambda_2}}$, and the equivalent MRC system with
respective Green's matrices ${\bf{C_1}}$ and ${\bf{C_2}}$. The
high accuracy of our approach is maintained for arbitrary branch
correlations.

\section{Conclusions}
A tight closed-form approximation to the distribution of the sum
of correlated Nakagami-$m$ RVs was introduced for the case of
identical and integer fading parameters. The proposed method
approximates this distribution by using the statistics of the
square-root of the sum of statistically independent Gamma RVs.
Examples indicate that the new approximation is highly accurate
over the entire range of abscissas. To demonstrate this more
rigorously, the proposed distribution is tested against the
computer generated data by the use of the Chi-square and the
Kologorov-Smirnov goodness-of-fit tests. In case of maximal
correlation, the proposed distribution becomes the exact
distribution.

The presented approach allowed to successfully tackle the famous
problem of analytical performance evaluation of an EGC system with
arbitrarily correlated and unbalanced Nakagami-$m$ branches. The
significance of the presented results is underpinned by the
existence of a large body of literature dealing with MRC
performance analysis, which permits highly accurate and efficient
EGC performance evaluation.

\appendices
\section{}
\renewcommand{\theequation}{\thesection.\arabic{equation}}
\setcounter{equation}{0} Using [\ref{r17}, Eqs. (9.212 (1)) and
(7.622 (1))], one has the following identity
\begin{eqnarray*}\label{b3}
J(m,a,p,q)= \frac{1}{\Gamma(m)} \quad \qquad \qquad \qquad \qquad \qquad \qquad \qquad \quad \nonumber \\
\times \int_0^\infty u^{m-1}e^{-u}
{}_1F_1\left(-\frac{p}{2};m;-au\right)
{}_1F_1\left(-\frac{q}{2};m;-au\right)du
\end{eqnarray*}
\begin{equation}
=(1+a)^{\frac{p}{2}}\left(\frac{1+2a}{1+a}\right)^{\frac{q}{2}}
{}_2F_1\left(m+\frac{p}{2},-\frac{q}{2};m;-\frac{a^2}{1+2a}\right)
\end{equation}
Using [\ref{r17}, Eq. (9.212 (3)), pp. 1023] with some simple
algebraic manipulations, the general form (\ref{W}) of the
coefficient $W(2,1,1)$ can be simplified as
\begin{eqnarray*}\label{b7}
W(2,1,1) = m_z\left(\frac{\Gamma(m_z+1/2)}{\Gamma(m_z)}\right)^2
\left(\frac{1-\sqrt\rho}{1+(N-1)\sqrt\rho}\right)^{m_z} \nonumber \\
\times \bigg[J(m_z,a,1,1) + \frac{a(m_z+1/2)^2}{m_z^2} \qquad
\qquad \qquad \qquad \quad
\nonumber\\
\times \, J(m_z+1,a,1,1) + \frac{a}{4 m_z^2}J(m_z+1,a,-1,-1)
\nonumber \\
-\frac{a(m_z+1/2)^2}{m_z^2}J(m_z+1,a,-1,1)\bigg]
\end{eqnarray*}
\begin{equation}
\end{equation}
where $a=\sqrt\rho/(1+(N-1)\sqrt\rho)$

\section{}
\setcounter{equation}{0} The unknown joint moments in
(\ref{4moment}), $E[Z_m^2 Z_i Z_j]$, $E[Z_m Z_i^2 Z_j]$, $E[Z_m
Z_i Z_j^2]$ and $E[Z_m Z_n Z_i Z_j]$ can be calculated from
[\ref{r10}, Eqs. (11) and (12)]. Here we derive their simpler and
computationally more efficient alternatives. The alternative to
[\ref{r10}, Eq. (12)] is derived directly from the definition of
the joint moment $E[Z_1 Z_2 Z_3 Z_4]$,
\begin{eqnarray} \label{c1_old}
E[Z_m Z_n Z_i Z_j]
=\int_0^\infty\int_0^\infty\int_0^\infty\int_0^\infty z_mz_nz_iz_j
\qquad \nonumber
\\
\times f_{Z_mZ_nZ_iZ_j}(z_m,z_n,z_i,z_j)dz_mdz_ndz_idz_j \,,
\end{eqnarray}
where joint pdf of four exponentially correlated Nakagami-$m$ RVs
is expressed as [\ref{r10}, Eq. (9)]
\begin{eqnarray}
f_{Z_mZ_nZ_iZ_j}(z_m,z_n,z_i,z_j) \qquad \qquad \qquad \qquad \qquad \quad \nonumber \\
=\frac{2^4 m_z^{m_z+3} |{\bf{\Psi}}|^{m_z}} {\Gamma(m_z)}
\frac{z_m^{m_z}z_nz_iz_j^{m_z}}{|\psi_{12} \psi_{23}
\psi_{34}|^{m_z-1}} \qquad \qquad \quad \nonumber\\
\times I_{m_z-1}\left(2m_z |\psi_{12}|z_m z_n\right)
I_{m_z-1}\left(2m_z |\psi_{23}|z_n z_i\right) \nonumber \\
\times I_{m_z-1}(2m_z|\psi_{34}|z_i z_j)
\, \exp\Big(-m_z\big(\psi_{11}z_m^2 \qquad \nonumber \\
+ \psi_{22}z_n^2 + \psi_{33}z_i^2 + \psi_{44}z_j^2\big)\Big),
\end{eqnarray}
where ${\bf{\Psi}} = [\psi_{i,j}]$ is defined by
(\ref{Psi_definition}). Now, we integrate [\ref{r10}, Eq. (11)]
over $z_m$ and $z_j$, respectively obtaining
\begin{eqnarray}
\int_0^\infty z_m^{m_z+1} \exp\left(-m_z \psi_{11} z_m^2\right) I_{m_z-1}\left(2 m_z \psi_{12}  z_m z_n\right) \nonumber \\
=\frac{1}{2}\left(m_z \psi_{12}z_n \right)^{m_z-1} \left(m_z
\psi_{11}\right)^{-(m_z+1/2)} \qquad \qquad \nonumber \\
\times \frac{\Gamma(m_z+1/2)}{\Gamma(m_z)}
{}_1F_1\left(m_z+\frac12; m_z;
m_z\frac{\psi_{12}^2}{\psi_{11}}z_n^2\right) ,
\end{eqnarray}
and
\begin{eqnarray}
\int_0^\infty z_j^{m_z+1} \exp\left(-m_z \psi_{44} z_j^2\right) I_{m_z-1}\left(2 m_z \psi_{34}  z_i z_j\right) \nonumber \\
=\frac{1}{2}\left(m_z \psi_{34} z_n\right)^{m_z-1} \left(m_z
\psi_{44}\right)^{-(m_z+1/2)} \qquad \qquad
\nonumber\\
\times \frac{\Gamma(m_z+1/2)}{\Gamma(m_z)}
{}_1F_1\left(m_z+\frac12; m_z;
m_z\frac{\psi_{34}^2}{\psi_{44}}z_i^2\right) \,.
\end{eqnarray}
We then use the series expansion of the modified Bessel function
of first kind [\ref{r17}, Eq. (8.445)] that allows to separate the
integrations per variables $z_n$ and $z_i$, yielding
(\ref{cetiri}). A similar procedure yields to an alternative of
[\ref{r10}, Eq. (11)] given by (\ref{tri}).

\section*{Acknowledgment}
The authors would like to thank the Editor and the anonymous
reviewers for their valuable comments that considerably improved
the quality of this paper.

\end{document}